\documentclass[aps,prl,superscriptaddress,twocolumn,longbibliography]{revtex4-1}
\usepackage{graphicx}
\usepackage{gensymb}
\usepackage{soul}
\usepackage[intlimits]{amsmath}
\usepackage{natbib,hyperref}
\hypersetup{colorlinks=true, citecolor=black, urlcolor=black, linkcolor=black}
\begin{document}
\title{Transverse Kerker Scattering for \AA{}ngstr\"o{}m Localization of Nanoparticles}
\author{Ankan Bag}
\author{Martin Neugebauer}
\author{Pawe{\l} Wo{\'z}niak}
\author{Gerd Leuchs}
\author{Peter Banzer}
\email{peter.banzer@mpl.mpg.de}
\affiliation{Max Planck Institute for the Science of Light, Staudtstr.\,2, D-91058 Erlangen, Germany}
\affiliation{Institute of Optics, Information and Photonics, Department of Physics, Friedrich-Alexander-University Erlangen-Nuremberg, Staudtstr.\,7/B2, D-91058 Erlangen, Germany}
%\affiliation{Department of Physics, University of Ottawa, 25 Templeton, Ottawa, Ontario, K1N 6N5 Canada}
\date{\today}
\begin{abstract}
\AA{}ngstr\"o{}m precision localization of a single nanoantenna is a crucial step towards advanced nanometrology, medicine and biophysics. Here, we show that single nanoantenna displacements down to few \AA{}ngstr\"o{}ms can be resolved with sub-\AA{}ngstr\"o{}m precision using an all-optical method. We utilize the tranverse Kerker scattering scheme where a carefully structured light beam excites a combination of multipolar modes inside a dielectric nanoantenna, which then upon interference, scatters directionally into the far-field. We spectrally tune our scheme such that it is most sensitive to the change in directional scattering per nanoantenna displacement. Finally, we experimentally show that antenna displacement down to 3 \AA{} is resolvable with a localization precision of 0.6 \AA{}.
\end{abstract}
\maketitle
%\section{\textbf{Introduction}}
Nanoantennas are fundamental building blocks in modern nanophotonic devices and experimental schemes. Accordingly, depending on the actual application, various antenna designs have been proposed and investigated in recent years~\cite{Taminiau2008,Bharadwaj2009,Giannini2011,Krasnok2013,Rybin2013}. In bio-sensing applications, bow-tie metal antennas can be used to substantially enhance the field locally, thus enabling single molecule detection and surface enhanced Raman spectroscopy~\cite{Kinkhabwala2009,Hatab2010}. On the other hand, Yagi-Uda type antennas are important components in optical circuitry to realize far-field to near-field coupling and vice-versa~\cite{Curto2010,Kriesch2013}. Furthermore, the spectral composition of an excitation field can be deduced from the pattern of the light scattered off a bi-metallic nanoantenna~\cite{Shegai2011}.

Besides the aforementioned complex antenna designs, even single-element antennas such as cylinders and spheres can exhibit interesting scattering properties, such as, directional emission and coupling via polarization dependent spin-momentum locking~\cite{Rodriguez-Fortuno2013,Neugebauer2014,Bliokh2015q}. Another example are scatterers, which fulfill Kerker's condition~\cite{Alaee2015,Kerker1983,Garcia-Camara2011,Nieto-Vesperinas2011}. This effect is typically associated with simultaneous excitation of electric and magnetic dipoles (Huygens' dipole), resulting in enhanced or suppressed forward/backward scattering~\cite{Geffrin2012,Person2013,Staude2017,Xi2016}. In Refs.~\cite{Neugebauer2014,Neugebauer2016,Xi2017,Wang2018}, a localization scheme was proposed and discussed based on enhanced or suppressed scattering in the direction transverse to propagation. Here, we present a detailed spectral analysis of transverse Kerker scattering off a single silicon particle. First, we elaborate on the general concept for a particle in free-space. Then, we present an actual experimental implementation and tune the wavelength of the excitation field to optimize the position-dependent directional scattering. By choosing optimal parameters, nanoantenna displacements down to few \AA{}ngstr\"o{}m can be resolved with sub-\AA{}ngstr\"o{}m precision and accuracy.
%\section{\textbf{Theory}}
\begin{figure}[h]
	%\hspace{-0.1in}
	\includegraphics[scale=1]{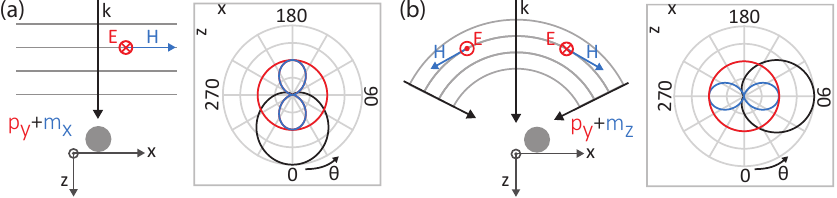}
	\caption{Kerker scattering and Huygens' dipole. (a) Zero backscattering. A $y$-polarized plane wave with $\mathbf{k}=k_z\hat{\mathbf{e}}_z$ excites transverse electric and magnetic dipoles $p_y$ and $m_x$ of equal amplitudes and phases, resulting in a Huygens' dipole and zero backscattering. (Inset) Polar plot of the resulting total far-field intensity of the scattered light (black) and the individual emission patterns of $p_y$ (red) and $m_x$ (blue) in the meridional $xz$-plane. (b) Directional scattering perpendicular to the propagation direction of the beam (transverse Kerker scattering). A tightly focused azimuthally polarized beam can induce longitudinal magnetic $m_z$ and transverse electric $p_y$ dipoles, resulting in transverse Kerker scattering [inset similar to (a)].}
	\label{kerker}
\end{figure}

As proposed by Kerker et al.~\cite{Kerker1983}, a plane-wave-like excitation of a small spherical particle ($\text{radius} \ll \lambda$) can result in asymmetric forward/backward scattering depending on the amplitudes and phases of the induced electric and magnetic dipole moments. A Huygens' dipole with zero back-scattering can be achieved when the induced electric and magnetic dipolar scattering coefficients are equal in amplitude and phase~\cite{Lukyanchuk2015,Paniagua-Dominguez2016}, as can be seen in the sketch in Fig.~\ref{kerker}(a). Equivalently, upon interference of a longitudinal and a transverse dipole, it is possible to achieve what we refer to as transverse Kerker scattering, that is, directional scattering perpendicular to the propagation direction of the excitation beam. Longitudinal particle modes can be excited by carefully structuring a three dimensional excitation field. For example, by inducing a combination of longitudinal magnetic and transverse electric dipoles with equal amplitude and phase, we obtain a similar emission pattern as in Fig.~\ref{kerker}(a), however rotated by $90\degree$, as sketched in Fig.~\ref{kerker}(b).
\begin{figure*}[t]
	%\hspace{-0.1in}
	\centering
	\includegraphics[scale=1]{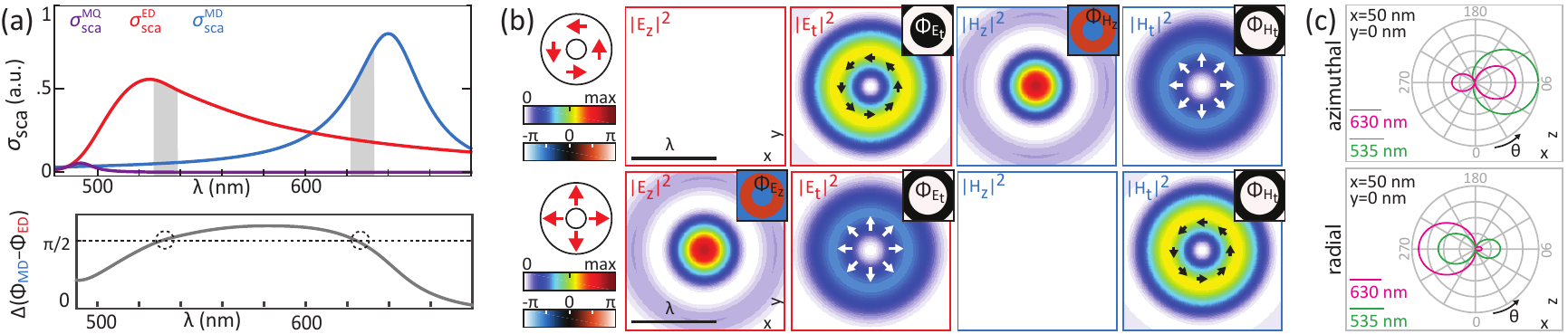}
	\caption{(a) Scattering cross-section $\sigma_{\text{sca}}$ of a silicon nanoantenna ($156$\,nm core diameter with $6$\,nm oxide shell) in free-space calculated via Mie theory. The nanoantenna predominantly supports magnetic (MD) and electric dipole (ED) modes along with a much weaker magnetic quadrupole (MQ) mode. The relative phase between MD and ED scattering coefficients is shown below. Around $\lambda\approx535$\,nm and $\lambda\approx630$\,nm, indicated by dotted black circles and corresponding gray regions, ED and MD scattering coefficients are $\pi/2$ out-of-phase. (b) Theoretical field intensity and relative phase (inset) distributions for tightly focused azimuthally (top) and radially (bottom) polarized beams. For an azimuthally (radially) polarized beam, the transverse electric field $\mathbf{E}_t$ ($E_x \hat{\mathbf{e}}_x + E_y \hat{\mathbf{e}}_y$) is azimuthally (radially) polarized, the transverse magnetic field $\mathbf{H}_t$ ($H_x \hat{\mathbf{e}}_x + H_y \hat{\mathbf{e}}_y$) is radially (azimuthally) polarized, and the longitudinal field component is purely magnetic (electric), with $E_z=0$ ($H_z=0$). All intensity plots are normalized to the maximum value of the total field intensity $I_{\text{tot}}=\left|\mathbf{E}\right|^2+\left|\mathbf{H}\right|^2$ (Gaussian units). (c) Polar plots of the resulting far-field intensities of the scattered light in the meridional $xz$-plane for a nanoantenna positioned at $\left(x,y\right)=\left(50,0\right)$\,nm in the focal plane. Transverse Kerker scattering is highly directional for azimuthal (radial) polarization at $\lambda_{\text{azim}}=535$\,nm ($\lambda_{\text{rad}}=630$\,nm) but weaker for $\lambda_{\text{azim}}=630$\,nm ($\lambda_{\text{rad}}=535$\,nm). %The corresponding other two plots show weaker directionality indicating a strong wavelength dependence of the directional scattering effect.
	}
	\label{beam&particle}
\end{figure*}

For the implementation of transverse Kerker scattering, high-refractive-index dielectric nanoparticles can be used, since they support electric and magnetic dipole modes of comparable strength ~\cite{Fu2013,GarciaEtxarri2011,Wozniak2015}. As an example, we consider a silicon nanoparticle with $156$\,nm core diameter and an estimated $6$\,nm silicon-dioxide shell, similar to the particle utilized in the experiment described below. The scattering off such a particle excited by a tightly focused beam can be treated by generalized Mie theory~\cite{Stratton1941}. In Fig.~\ref{beam&particle}(a), we plot the scattering cross-sections of the electric dipole (ED), magnetic dipole (MD), and magnetic quadrupole (MQ) as red, blue, and purple lines within the visible spectral range for the particle in free-space. In the dominant part of the depicted spectrum, the MQ contribution can be neglected and we can approximate the nanoparticle as a dipole, such that the induced dipole moments are proportional to the local electromagnetic field components, $\mathbf{p} \propto T_{\text{ED}}\mathbf{E}$ and $\mathbf{m} \propto T_{\text{MD}}\mathbf{H}$~\cite{Neugebauer2016}. The proportionality factors, $T_{\text{ED}}$ and $T_{\text{MD}}$ are the electric and magnetic dipole scattering coefficients calculated using Mie theory, which define the strength and phase of the induced dipole moments~\cite{Tsang2000}. They are linked to the scattering cross-sections by $\sigma_{\text{sca}}^{\text{ED}}\propto\left|T_{\text{ED}}\right|^{2}$ and $\sigma_{\text{sca}}^{\text{MD}}\propto\left|T_{\text{MD}}\right|^{2}$~\cite{Tsang2000}. Since achieving transverse Kerker scattering depends not only on amplitudes but also on phases, we depict the relative phase between the two dominant ED and MD resonances in the lower graph of Fig.~\ref{beam&particle}(a). The black dotted circles and corresponding gray areas around $\lambda=535$\,nm and $630$\,nm denote the wavelength ranges where electric and magnetic dipoles are $\pi/2$ out-of-phase. The importance of these wavelengths will become clear, when we discuss the impinging light field in the following.

For excitation, we use tailored inhomogeneous electromagnetic field distributions obtained by tightly focusing azimuthally and radially polarized vector beams~\cite{Quabis2000,Youngworth2000,Rubinsztein-Dunlop2017}. Fig.~\ref{beam&particle}(b) shows the focal-plane electric and magnetic intensity and phase distributions of the field components: $\mathbf{E}_z$, $\mathbf{E}_t$ ($E_x \hat{\mathbf{e}}_x + E_y \hat{\mathbf{e}}_y$), $\mathbf{H}_z$, and $\mathbf{H}_t$ ($H_x \hat{\mathbf{e}}_x + H_y \hat{\mathbf{e}}_y$), calculated using vectorial diffraction theory~\cite{Richards1959,L.NovotnyandB.Hecht2006}. Both, the intensity and phase distributions of the transverse and longitudinal field components exhibit cylindrical symmetry. The amplitudes of the transverse components $\left| \mathbf{E}_t \right|$ and $\left|\mathbf{H}_t\right|$ are zero on the optical axis for both input beams, and in close proximity to the optical axis ($r\ll \lambda$), can be approximated to increase linearly with radial distance~\cite{Neugebauer2016}. The intensity of the longitudinal field component---only $\mathbf{E}_z$ ($\mathbf{H}_z$) is present for radial (azimuthal) polarization---are maximum on the optical axis and significantly stronger than the transverse components for $r\ll\lambda$. Another important aspect is the phase retardation of $\pm\pi/2$ between the longitudinal and transverse field components [see insets in Fig.~\ref{beam&particle}(b)]. By choosing the excitation wavelengths $\lambda=535$\,nm and $630$\,nm, the aforementioned phase difference between $T_{\text{ED}}$ and $T_{\text{MD}}$ cancels the phase retardation between the longitudinal and transverse components of the excitation fields.

\begin{figure*}[htbp]
	\hspace*{-0.2in}
	\centering
	\includegraphics[scale=1.05]{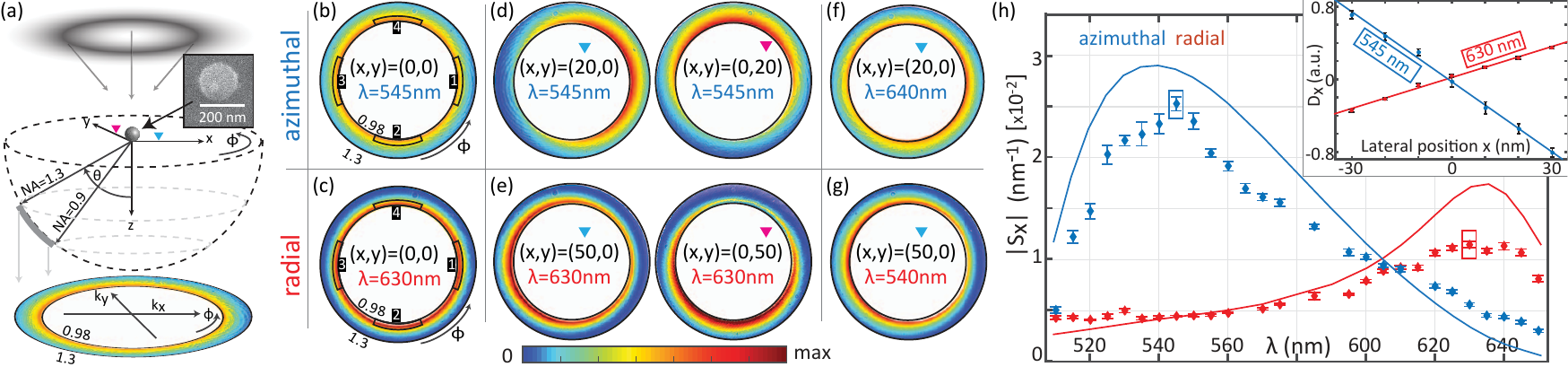}
	\caption{ Experimental implementation. (a) Tight-focusing of a collimated input beam (radially or azimuthally polarized) using a microscope objective of numerical aperture (NA) of 0.9 onto a silicon nanoantenna sitting on a glass substrate (SEM image in inset), which can be precisely positioned by a piezo-stage. The forward scattered light is collected using an oil-immersion type microscope objective of NA=1.3. The back-focal plane (BFP) intensity distribution (for NA$\in$[0.98,1.3]) is captured onto a CCD camera. (b-g) Exemplary BFP images showing symmetric scattering for an on-axis position (b-c), and transverse Kerker directionality for off-axis positions (d-g). The strength of directional scattering strongly depends on the wavelength as shown in (d-g) with maximum directionality observed at $\lambda_{\text{azim}}=545$\,nm (d) and $\lambda_{\text{rad}}=630$\,nm (e). (h) Sensitivity ($S_x$) or change in directivity $D_x$ per nm---$\left| S_x \right|$ is plotted for all wavelengths along with the numerically calculated results for comparison (radial and azimuthal polarization in red and blue). (Inset) Plot of the measured directivity $D_x$ versus the antenna's $x$-position for $\lambda_{\text{rad}}=630$\,nm (in red) and $\lambda_{\text{azim}}=545$\,nm (in blue) where $\left|S_x\right|$ is strongest.}
	\label{setup&results}
\end{figure*}

Consequently, when the wavelength is optimized with respect to the resonances of the particle, longitudinal and transverse dipole moments with a relative phase of $0$ or $\pi$ can be induced, resulting in transverse Kerker scattering. Furthermore, the relative amplitudes between longitudinal and transverse dipole moments can be adapted by changing the radial distance between the particle and the optical axis. Therefore, our system enables tailoring of transverse Kerker scattering, which will be discussed with examples. The nanoantenna positioned in the focal plane at $\left(x,y\right)=\left(50,0\right)$\,nm can be considered as a combination of $m_z$, $p_y$, and $m_x$ dipoles for the azimuthally polarized beam, and $p_z$, $m_y$, and $p_x$ dipoles for the radial one. The relative amplitudes and phases of each dipole moment can be determined from Figs.~\ref{beam&particle}(a)~and~(b). In Fig.~\ref{beam&particle}(c), we plot the resulting far-field intensity of the scattered light in the meridional $xz$-plane for $\lambda=535$\,nm (green) and $\lambda=630$\,nm (magenta). Highly directional transverse Kerker scattering can be observed for azimuthal polarization at $\lambda_{\text{azim}}=535$\,nm (directionality in positive $x$-direction) and for radial polarization at $\lambda_{\text{rad}}=630$\,nm (directionality in negative $x$-direction). In contrast, the other two corresponding plots indicate a much weaker directionality, highlighting the wavelength dependence of the transverse Kerker scattering, which will be discussed in detail below. 

In the actual experimental implementation of transverse Kerker scattering based localization, the nanoantenna is placed on a dielectric interface (air-glass), which substantially modifies the scattering scheme from the aforementioned free-space scenario. To analytically describe the full scattering process, we start with determining the complete scattering matrix $\hat{\mathbf{T}}$ of the nanoantenna sitting on an interface~\cite{Mishchenko2002}, such that the incident field $\mathbf{E}^{\text{inc}}$ and the scattered field $\mathbf{E}^{\text{sca}}$ are related by $\mathbf{E}^{\text{sca}} = \hat{\mathbf{T}} \mathbf{E}^{\text{inc}}$. We expand our highly confined focal field into electromagnetic multiploes~\cite{Orlov2012,Mojarad2008,Hoang2012} as
\begin{equation}
	\mathbf{E}^i = \sum^{\infty}_{n=1} \sum^{n}_{m=-n} a^i_{mn} \mathbf{N}^i_{mn} + b^i_{mn} \mathbf{M}^i_{mn}\text{,}
	\label{eqnVSH}
\end{equation}
where $i$ corresponds to either the incident or the scattered fields. $\mathbf{N}^i_{mn}$ and $\mathbf{M}^i_{mn}$ are vector spherical harmonics (regular or irregular type for incident or scattered field respectively) representing the electric and magnetic multipoles expanded around the center of the particle~\cite{Orlov2012}. The complex-valued multipole expansion coefficients $a^{\text{inc}}_{mn}$ and $b^{\text{inc}}_{mn}$ for the incident field contain full information about the electric $\mathbf{E}^{\text{inc}}(\mathbf{r})$ and magnetic $\mathbf{H}^{\text{inc}}(\mathbf{r})$ field components at each point $\mathbf{r}$. Following ~\cite{Tsang2000,Bauer2013}, we model the influence of the interface by considering the effect of incident and scattered light reflected from the interface. Hence, the expansion coefficients $a^{\text{sca}'}_{mn}$ and $b^{\text{sca}'}_{mn}$ representing the induced multipole moments can be obtained from the auxiliary scattered field above the interface $\mathbf{E}^{\text{sca}'}$, which is related to $\mathbf{E}^{\text{inc}}$ via the effective scattering matrix $\hat{\mathbf{T}}_{\text{eff}}$ as
\begin{equation}
	\mathbf{E}^{\text{sca}'} = \hat{\mathbf{T}}_{\text{eff}} \mathbf{E}^{\text{inc}} = \frac{ \hat{\mathbf{T}} (1+\mathbf{L}^{(1)}_R) }{ 1-\hat{\mathbf{T}} \mathbf{L}^{(3)}_R} \mathbf{E}^{\text{inc}}
	\label{eqnTeff}
\end{equation}
where $\mathbf{L}^{(1,3)}_R$ are the reflection operators of the interface for the incident and scattered light (see more details in the supplementary of ~\cite{Bauer2013}). These expansion coefficients $a^{\text{sca}'}_{mn}$ and $b^{\text{sca}'}_{mn}$ can then be used to calculate the light emitted into the glass substrate $\mathbf{E}^{\text{sca}}$ taking into account the transmission Fresnel coefficients~\cite{Orlov2012,Bauer2013}. In particular, we consider the peak emission at the critical angle, which is equivalent to the transverse plane in free-space~\cite{L.NovotnyandB.Hecht2006}. The transmitted far-field intensity $I \propto \left|\mathbf{E}^{\text{sca}}_{\text{far}}\right|^2 $ along the critical angle is then used to numerically calculate the wavelength and position-dependent transverse Kerker scattering for azimuthally and radially polarized beams~\footnote{See Supplementary Material for more details about the theoretical calculation, which includes Ref.~\cite{Cruzan1961}; and for more details about the experimental set-up, data acquisition, processing, and error propagation, which includes Ref.~\cite{Marrucci2006}}.

%\section{\textbf{Experiment}}
Regarding the experimental implementation, Fig.~\ref{setup&results}(a) depicts a schematic sketch where we tightly focus an incoming beam with a microscope objective of numerical aperture (NA) of 0.9 onto a silicon nanoantenna (inset) sitting on a glass substrate, which can be precisely positioned within the focus by a piezo-stage. We collect the transmitted and forward scattered light with a second (oil-immersion type) microscope objective of NA=1.3, and image the angularly resolved intensity distribution $\tilde{I}(k_x,k_y)$ of the back-focal plane (BFP) of said objective onto a CCD camera~\cite{Note1}. Similar to Ref.~\cite{Neugebauer2016}, we only consider the region, NA$\in$[0.98,1.3], where we can detect the scattered light without the transmitted beam. For each wavelength, we scan our nanoantenna within the focal plane and obtain BFP images (exposure time ~1\,ms) for each ($x,y$) position [examples plotted in Figs.~\ref{setup&results}(b-g)].

To define the position-dependent strength of the directional scattering, we calculate the difference between the light scattered into opposite directions in $k$-space ($k_x,k_y$): $D_x=(I_3-I_1)/I_{\text{tot}}$ and $D_y=(I_2-I_4)/I_{\text{tot}}$, where $I_{\text{tot}}=\frac{1}{2}\Sigma_{j=1}^4 I_j$. Here, $I_j$ ($j\in[1,4]$) is the average intensity of the $j^{th}$ region in the BFP, which corresponds to the angular regions defined by $\Delta \phi = 45 \degree$ around $\pm k_x$ and $\pm k_y$ and NA$\in$[0.98,1.03] as indicated in Figs.~\ref{setup&results}(b-c). The choice of these angular regions is due to a stronger and uniform scattering signal~\cite{Note1}. Figs.~\ref{setup&results}(d-g) show exemplary BFP images indicating varying extends of transverse directivity for different wavelengths. For each wavelength, we obtain $D_x(x)$ and $D_y(y)$ curves by fitting our experimentally measured directivity for the nanoantenna displacement along the $x$- and $y$-axis. In line with the behavior of the transverse electromagnetic fields, the direcitivity $D_x(x)$ and $D_y(y)$ exhibit a linear relationship with displacement within at least $30$\,nm around the optical axis. In inset of Fig.~\ref{setup&results}(h), without loss of generality, $D_x(x)$ is plotted against the $x$-position of the particle for azimuthally and radially polarized beams of wavelengths $\lambda_{\text{azim}}=545$\,nm and $\lambda_{\text{rad}}=630$\,nm. Following the chosen definition of directivity $D_{x(y)}$, it can be seen that the slope of $D_x$ is negative for azimuthally polarized beam, as can also be observed in the exemplary BFP images in Figs.~\ref{setup&results}(b,d,f), when compared to Figs.~\ref{setup&results}(c,e,g) for radial polarization. Each data point in these two calibration curves is a statistical representation of more than $50$ measurement values. Fitting fluctuations shown as error bars represent the stability of our current experimental setup and do not reflect upon the localization resolution (see more details below). As we can see, a very high directivity of $ D_x \approx 0.35$ for radial and even higher directivity of $ D_x \approx 0.76$ for azimuthal polarization can be achieved for a displacement as small as $30$\,nm. 

In order to quantify the position sensitivity of our experiment, we define the parameter sensitivity $S(x,y)$ as average change in directivity along $x$- and $y$-axis within the region of linearity, $S_x=\partial D_x/ \partial x$, $S_y=\partial D_y/ \partial y$. With respect to plots in the inset of Fig.~\ref{setup&results}(h), $S_x$ represents the slopes of the two linear fits. In Fig.~\ref{setup&results}(h), we present a spectral analysis of $\left| S_x \right| \pm \Delta S_x$ for radial (red) and azimuthally polarized (blue) beams where $\Delta S_x$ is the fitting error of the slope represented as errorbars. For comparison, we plot the numerically calculated results (bold lines) based on the theoretical model presented earlier. The parameter $\left| S_x \right|$ is maximum around $\lambda_{\text{azim}}=545$\,nm and $\lambda_{\text{rad}}=630$\,nm, where the electric and magnetic dipoles are induced with a relative phase close to $0$ or $\pi$ [see free-space scenario in Fig.~\ref{beam&particle}(a)]. Also, for azimuthal polarization ($\lambda_{\text{azim}}\approx545$\,nm), $\left|S_{x(y)} \right|$ is more than two times stronger than for radial polarization ($\lambda_{\text{rad}}\approx630$\,nm). The result highlights that, for the utilized silicon particle, the best choice for transverse Kerker scattering based localization is an azimuthally polarized excitation beam with $\lambda_{\text{azim}}=545$\,nm, since a stronger directivity leads to enhanced localization accuracy.

%\section{\textbf{Angstrom localization}}
\begin{figure}[t]
	\hspace*{-0.1in}
	\centering
	\includegraphics[scale=0.85]{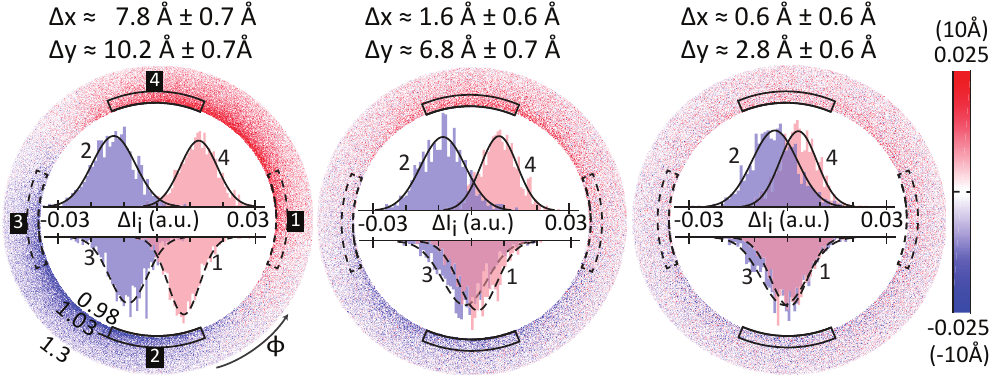}
	\caption{Differential BFP images $\Delta \tilde{I}$ for relative displacements of less than $1$\,nm for an azimuthally polarized excitation beam with $\lambda_{\text{azim}}=545$\,nm ($S_x \approx S_y \approx -0.025\,\text{nm}^{-1}$). In each BFP image, $\Delta \tilde{I_j}$ values from the outlined regions are used to plot the histograms shown in the centers; dash-outlined regions ($j=1,3$) correspond to $\Delta x$ and solid ones ($j=2,4$) to $\Delta y$. Gaussian fits to the $\Delta \tilde{I_j}$ distributions (histogram) are used to estimate the relative displacements (peak-to-peak) and the precision of estimation (standard error of the mean). The Gaussian peaks are clearly distinguishable for a relative displacement down to $3$\,\AA{}, with precision of $0.6$\,\AA{}.}
	\label{angstrom}
\end{figure}
To demonstrate our best localization accuracy, we resolve relative displacements of the nanoparticle by considering differential BFP images---difference between two BFP intensity distributions corresponding to two particle positions, $\Delta \tilde{I}=[\tilde{I}(x_1,y_1)-\tilde{I}(x_2,y_2)]/I_{\text{tot}}$---for $\lambda_{\text{azim}}=545$\,nm, where the relative nanoantenna displacements were less than $1$\,nm. This way we can obtain the displacement of the particle between two subsequently recorded positions and, hence, also the location of the particle with respect to the optical axis. However, achieving such small displacements was not possible deterministically with our setup ($\pm$4\,nm position-inaccuracy). Therefore, we raster scanned our nanoparticle with 2\,nm step size within $x=[-20,20]$ nm and $y=[-20,20]$ nm around the optical axis (region of linearity) and captured BFP images with 1\,ms exposure time for each position. This leads to a collection of position pairs, some being only few Angstroms apart, some others nanometers apart, owing to the positioning inaccuracy of our system. Exemplary differential BFP images $\Delta \tilde{I}$ are shown in Fig.~\ref{angstrom} where we once again consider the four angular regions $j$, $j\in[1,4]$ defined by $\Delta \phi = 45 \degree$ and NA$\in$[0.98,1.03]. The central histogram plots shows pixel-intensity distributions of $\Delta \tilde{I_j}$, for the $j^{\text{th}}$ regions marked in the differential BFP images. Region $j=1,3$ (dashed black border) corresponds to a movement $\Delta x$ and region $j=2,4$ (solid black border) for $\Delta y$. Gaussian fits to the histograms of $\Delta \tilde{I_j}$ allow us to estimate the relative displacements along $x$- and $y$-axis such that 
\begin{equation}
\Delta x = \frac{\Delta I_3 - \Delta I_1}{S_x}, \Delta y = \frac{\Delta I_2 - \Delta I_4}{S_y}.
\label{eqnresolve}
\end{equation}
where $\Delta I_j$ represents the expectation value of the pixel-wise distribution of $\Delta \tilde{I_j}$. The localization precision is calculated using an error propagation formula considering both $\Delta S_{x(y)}$ and $\sigma_{M,j}$, with $\sigma_{M,j}=\sigma_j / \sqrt{\text{\#pixels in } j^{\text{th}} \text{ region}}$ being the standard error of the mean of the fitted Gaussian~\cite{Note1}. Fig.~\ref{angstrom} shows clearly distinguishable Gaussian peaks for a displacement down to $3$\,\AA{} with precision of $0.6$\,\AA{}, whereas, for less than $3$\,\AA{} displacements, the Gaussian peaks overlap significantly with our current experimental setup.

%\section{\textbf{conclusion}}
In conclusion, we first discuss theoretically a simple transverse Kerker scattering scheme in free-space, consisting of a tightly focused vector beam and a spherical dielectric nanoparticle. This scheme might find practical application in tweezers systems. Next, we extended the scheme for a more sophisticated experimental scenario (particle-on-interface) with an analytical model, which can be applied to arbitrary excitation beams and particle parameters, such as size, refractive index etc. Moreover, the effect of parameter variation can be taken into account to optimize the directional scattering. Finally, our experimental results, which are in good agreement with the analytical results, show that, upon optimization, an individual nanoantenna displacement down to few \AA{}ngstr\"o{}m can be resolved with sub-\AA{}ngstr\"o{}m accuracy. The discussed scheme proves that the location of nanoparticles can be sensed with ultra-high precision and accuracy, paving the way towards interesting applications, such as, stabilization of positioning systems in microscopy and nanometrology. Moreover, a quadrant-detector based signal detection would allow for an ultra-fast time-resolved tracking of nanoscopic systems.\\
\\
\section*{Acknowledgment}
We thank Thomas Bauer for the fruitful discussions.

\bibliography{bib}
\end{document}